\documentstyle[12pt]{article}
\input{psfig}
\long\def\comment#1{}
\begin{document}
\title{The Solar Numbers in Angkor Wat}

\author{Subhash Kak\\
Department of Electrical \& Computer Engineering\\
Louisiana State University\\
Baton Rouge, LA 70803-5901, USA\\
FAX: 225.388.5200; Email: {\tt kak@ee.lsu.edu}}

\maketitle

\begin{abstract}
The great Vi\d{s}\d{n}u temple at Angkor Wat in north-central Kampuchea
(Cambodia)
is known to have been built according to an astronomical plan.
In this note we show that the little-understood solar formula of the temple is
identical to the one in the {\it \'{S}atapatha Br\={a}hma\d{n}a.}
We propose that the Angkor Wat formula was an expression
of the 
{\it \'{S}atapatha} astronomy.

\end{abstract}

The great Vi\d{s}\d{n}u temple
of Angkor Wat was built by the Khmer Emperor
S\={u}ryavarman II, who reigned during AD 1113-50.
This temple was one of
the many temples built from AD 879 - 1191, when the
Khmer civilization
was at the height of its power.
The Vi\d{s}\d{n}u temple
has been called one of humankind's most impressive and enduring
architectural achievements. 

More than 20 years ago {\it Science} carried a comprehensive
analysis by Stencel, 
Gifford and Mor\'{o}n (SGM) of the astronomy and cosmology underlying
the design of this temple.$^1$
The authors concluded that it served as a practical
observatory where the rising sun was aligned on the equinox
and solstice days with the western entrance of the temple,
and they identified 22 sighting lines for seasonally 
observing the risings of the sun and the moon.
Using a survey by Nafilyan$^2$ and converting the figures
to the Cambodian cubit or {\it hat} (0.435 m), SGM demonstrated
that certain 
measurements of the temple
record calendric and cosmological time cycles.

In addition, SGM showed that the west-east axis 
represents the periods of the yugas.
The width of the moat is 439.78 {\it hat};
the distance from the first step of the western
entrance gateway to balustrade wall at the end of
causeway is 867.03 {\it hat};
the distance from the first step of the western
entrance gateway to the first step of the central tower
is 1,296.07 {\it hat}; and the distance from the first step of
bridge to the geographic center of the temple is 1,734.41
{\it hat}. These correspond to the periods of
432,000; 864,000; 1,296,000; 1,728,000 years for
the Kali, Dv\={a}para, Tret\={a}, and K\d{r}ta yuga,
respectively. SGM suggest that the very slight discrepancy
in the equations might be due to human error or 
erosion or sinking of the structure.

In the central tower, the topmost elevation has external
axial dimensions of 189.00 {\it hat} east-west, and
176.37 {\it hat} north-south, with the sum of 365.37.
In the words of SGM, this is ``perhaps the most
outstanding number'' in the complex, ``almost
the exact length of the solar year.''
But SGM were not able to explain the inequality of
the two halves; which is the problem that
we take up in this paper.
We will show that these numbers are old
{\it \'{S}atapatha Br\={a}hma\d{n}a} numbers for
the asymmetric motion of the sun.

\section*{The Historical Background of Angkor Wat}

The kings of the Khmer empire ruled over a vast
domain that reached from what is now southern Vietnam to
Yunan, China and from Vietnam westward to the Bay of Bengal.
The structures one sees at Angkor today, more than 100
temples in all, are the surviving religious remains of a
grand social and administrative metropolis whose other
buildings - palaces, public buildings, and houses - were all
built of wood and are long since decayed and gone.
As in most parts of India where wood
was plentiful, only the gods had the right
to live in houses of stone or brick; the sovereigns 
and the common folk lived in pavilions and houses of wood.

Over the
half-millenia of Khmer rule, the city of Angkor became
a great pilgrimage destination because of the notion of
Devar\={a}ja, that 
has been explained by
Lokesh Chandra as
a coronation icon. 
Jayavarman II
(802-850) was the first to use this royal icon.
According to Lokesh Chandra,
\begin{quotation}
Devar\={a}ja means `King of the Gods' and
not `God-King'. He is Indra and refers to the
highly efficacious {\it aindra} mah\={a}bhi\d{s}eka
of the \d{R}gvedic r\={a}jas\={u}ya tradition as
elaborated in the Aitareya-br\={a}hma\d{n}a.
It was not a simple but a great coronation,
a mah\={a}bhi\d{s}eka. It was of extraordinary significance
that Jayavarman II performed a \d{R}gvedic rite, which lent
him charismatic authority.$^3$
\end{quotation}

The increasingly larger temples built by the
Khmer kings continued to function as the locus of the devotion to the
Devar\={a}ja, and were at the same time earthly and symbolic
representations of mythical Mt. Meru, the cosmological home of
the Hindu gods and the axis of the
world-system. The symbol of the king's divine authority
was the sign ({\it linga}) of
\'{S}iva within the temple's inner sanctuary,
which represented both the axes of physical and the
psychological worlds.
The worship of \'{S}iva and 
Vi\d{s}\d{n}u separately, and together as Harihara,
had been popular for considerable
time in southeast Asia;
Jayavarman's chief innovation
was to use ancient Vedic mah\={a}bhi\d{s}eka to
define the symbol of government.
To quote Lokesh Chandra further,
``The icon used by Jayavarman II for his aindra
mah\={a}bhi\d{s}eka, his Devar\={a}ja = Indra (icon), became the
symbol of the Cambodian state, as the sacred and secular
sovereignty denoted by
Praj\={a}pat\={\i}\'{s}vara/Brahm\={a}, as the continuity
of the vital flow of the universal ({\it jagat}) into the
stability of the terrestrial kingdom ({\it r\={a}ja = r\={a}jya}).
As the founder of the new Kambuja state, he contributed a 
national palladium under its Cambodian appellation
{\it kamrate\.{n} jagat ta r\={a}ja/r\={a}jya}.
Whenver the capital was transferred by his successors,
it was taken to the new nagara, for it had to be constantly
in the capital.''$^4$

Angkor Wat is the supreme masterpiece of Khmer
art. 
The descriptions of the temple
fall far short of communicating the
great size, the perfect proportions, and the astoundingly
beautiful sculpture that everywhere presents itself to the
viewer. 

As an aside, it should be mentioned that 
some European scholars tended to date
Angkor Wat as being after the 14th century. The principal reason
was that some decorative motifs at Angkor Wat show
a striking resemblance to certain motifs of the
Italian Renaissance. This argument, which is similar
to the one used in dating Indian mathematical texts 
vis-a-vis Greek texts, has been proven to be
wrong. In the words of C\oe d\`{e}s,$^5$ ``If there is some
connexion between the twelfth-century art of the
Khmers, the direct heirs to the previous centuries,
and the art of the Renaissance, it must have been
due to a reverse process, that is to the importation of
oriental objects into Europe.''

\section*{Astronomy of Altars and Temples}

To understand the astronomical aspects of Angkor Wat
it is necessary to begin with the Indian
traditions of altar and temple design on which it
is based.
And since the Angkor Wat ritual hearkened to the 
Vedic past, it stands to reason that its
astronomy was also connected to the Vedic astronomical
tradition.

In a series of publications I have shown that
the Vedic altars had an astronomical basis.$^6$
In the basic scheme, the circle represented the
earth and the square represented the heavens
or the deity.
But the altar or the temple, as a representation of
the dynamism of the universe, required a
breaking of the symmetry of the square.
As seen clearly in the agnicayana and
other altar constructions, this was
done in a variety of ways.
Although the main altar might be square or
its derivative, the
overall sacred area was taken to be
a departure from this shape.
In particular, the temples to the goddess were
drawn on a rectangular plan.
In the introduction to the
{\it \'{S}ilpa Prak\={a}\'{s}a}, a 9th-12th century
Orissan temple architecture text, Alice Boner writes,$^7$
``[the Dev\={\i} temples]
represent the creative expanding forces, and therefore
could not be logically be represented by a square, which
is an
eminently static form. While the immanent supreme
principle is represented by the number ONE, the first stir
of creation initiates duality, which is the number TWO,
and is the producer of THREE and FOUR and all subsequent
numbers upto the infinite.''
The dynamism is expressed by a doubling of the square
to a rectangle or the ratio 1:2, where the garbhag\d{r}ha is now
built in the geometrical centre.
For a three-dimensional structure, the basic
symmetry-breaking ratio is 1:2:4, which can be continued further
to another doubling.$^8$

The constructions of the Harappan period (2600-1900 BC)
appear to be according to the same principles.
The dynamic ratio of 1:2:4 is the most commonly encountered
size of rooms of houses, in the overall plan of houses
and the construction of large public buildings.
This ratio is also reflected in the overall plan of the
large walled sector at Mohenjo-Daro called the citadel mound.
It is even the most commonly encountered
brick size.$^9$

There is evidence of temple structures in the Harappan
period in addition to iconography that recalls the
goddess. Structures dating to 2000 BC, built in the design of yantras,
have been unearthed in northern Afghanistan.$^{10}$
There is ample evidence for a continuity in the
religious and artistic tradition of India from
the Harappan times, if not earlier.
These ideas and
the astronomical basis continued in the
architecture of the temples of the classical
age.
Kramrisch has argued that the number 25,920, the number
of years in the precessional period of the earth, is also
reflected in the plan of the temple.$^{11}$

According to the art-historian Alice Boner,$^{12}$
\begin{quote}
[T]he temple must, in its space-directions, be
established in relation to the motion of the
heavenly bodies. But in asmuch  as it incorporates
in a single synthesis the unequal courses of the sun,
the moon and the planets, it also symbolizes
all recurrent time sequences: the day, the month,
the year and the wider cycles marked by the
recurrence of a complete cycle of eclipses, when the
sun and the moon are readjusted in their original 
positions, anew cycle of creation begins.
\end{quote}

It is clear then that the Hindu temple
is a conception of the astronomical
frame of the universe.
In this conception it serves the same
purpose as the Vedic altar, which served
to express the motions of the sun and the moon.
The progressive complexity of the classical
temple was inevitable given an attempt to
bring in the cycles of the planets and
other ideas of the yugas into
the scheme.

A text like the {\it \'{S}ilpa Prak\={a}\'{s}a} would
be expected to express the principles of 
temple construction of the times that led to
the Angkor Wat temple.
Given the prominence to the yuga periods in 
Angkor Wat and a variety of other evidence
it is clear that there
is a continuity between the Vedic and Pur\={a}\d{n}ic
astronomy and cosmology and the design of Angkor Wat.

\section*{Solar and lunar measurements}
Some of the solar and lunar numbers that
show up in the design of the Angkor Wat temple
are the number of nak\d{s}atras, the number
of months in the year, the days in the lunar month,
the days of the solar month, and so so.$^{13}$
Lunar observations appear to have been made from the
causeway.
SGM list 22 alignments in their paper, these could have
been used to track not just the solar and lunar motions but also
planetary motions.

The division of the year into the two halves:
189 and 176.37 has puzzled SGM.
But precisely the same division is described
in the {\it \'{S}atapatha Br\={a}hma\d{n}a}.
In layer 5 of the altar described in the 
\'{S}atapatha, a division of the year
into the two halves in the proportion
15:14 is given.$^{14}$
This proportion corresponds to the numbers
189 and 176.4 which are just the numbers used
at Angkor Wat.

Consider the physics behind the asymmetry
in the sun's orbit.
The period from the autumnal equinox to
the vernal equinox is smaller than the opposite circuit.
The interval between successive perihelia, the
anomalistic year, is 365.25964 days which is 0.01845 days
longer than the tropical year on which our calendar is based.
In 1000 calendar years, the date of the perihelion advances about
18 days.
The perihelion was roughly on December
18 
during the time of the construction of
Angkor Wat; and it was on October 27 
during early 2nd millennium BC, the most likely period of
the composition of the {\it \'{S}atapatha Br\={a}hma\d{n}a}.
In all these cases the perihelion occurs during the
autumn/winter period, and so by Kepler's 2nd law we know
that the speed of the sun in its orbit around the earth
is greater during the months autumn and winter than
in spring and summer.

During the time of the 
{\it \'{S}atapatha Br\={a}hma\d{n}a},
the apogee was about midway through the spring
season, which was then somewhat more than 94 days.
The extra brick in the spring quadrant may
symbolically reflect the discovery that this
quarter had more days in it, a discovery made at
a time when a satisfactory formula had not yet been developed
for the progress of the sun on the ecliptic.

It is possible that the
period from the spring equinox to the fall equinox was
taken to be about 189 days by doubling the
period of the spring season; 176 days became the period of the reverse
circuit.

Why not assume that there was no more to these
numbers than a division into the proportions
15:14 derived from some numerological
considerations?
First, we have the evidence from the
{\it \'{S}atapatha Br\={a}hma\d{n}a} that expressly informs
us that the count of days from the
winter to the summer solstice was different, and
shorter, than the count in the reverse order.
Second, the altar design is explicitly about
the sun's circuit around the earth and so the
proportion of 15:14 must be converted into
the appropriate count with respect to the length
of the year.
Furthermore, the many astronomical alignments of the
Angkor Wat impress on us the fairly elaborate
system of naked-eye observations that were the
basis of the temple astronomy.

But since precisely the same numbers were used
in Angkor Wat as were mentioned much earlier in
the {\it \'{S}atapatha Br\={a}hma\d{n}a}, one would presume that
these numbers were used as a part of ancient
sacred lore. 
We see
the count between the solstices has been
changing much faster than the count between the
equinoxes because the perigee has been,
in the past two thousand years somewhere between
the autumn and the winter months. 
Because of its relative constancy, the count between
the equinoxes became one of the primary `constants'
of Vedic/Pur\={a}\d{n}ic astronomy.

The equinoctial half-years are currently
about 186 and 179, respectively;
and were not much different when Angkor Wat temple
was constructed.
Given that the length of the year was known to considerable
precision there is no reason to assume that these
counts were not known.
But it appears that a `normative' division 
according to the ancient proportion was used.

As it was known that the solar year was about 365.25 days,
the old proportion of 15:14 would give the distribution
188.92 and 176.33, and that is very much the Angkor Wat
numbers of 189 and 176.37 within human error.
In other words, the choice of these `constants' may have been
determined by the use of the ancient proportion
of 15:14.

\section*{Conclusions}
It has long been known that the Angkor Wat temple astronomy
is derived from 
Pur\={a}\d{n}ic and Siddh\={a}ntic ideas.
Here we present 
evidence that takes us to 
the Vedic roots for the division of the solar year
in Angkor Wat
into two unequal halves.
This division is across
the equinoxes and that number has not changed very
much during the passage of time from the
{\it Br\={a}hma\d{n}as} to the construction
of the Angkor Wat temple, so it is not surprising
that it figures so prominently in the astronomy.
It also appears that the count of 189 days may have been
obtained by a doubling the measured period for
the spring season. 

The astronomy of Angkor Wat has the lesson that the
medieval and ancient Indian temple complexes, which
were also built with basic astronomical observations in mind,
 should
be examined for their astronomical bases.

\section*{Acknowledgement}
The author is thankful to F. Graham Millar of Halifax,
Nova Scotia for bringing the solar equation
in Angkor Wat to his attention and for useful discussions.

\section*{References}

\begin{enumerate}

\item Stencel, R.,  Gifford, F., Mor\'{o}n, E.,
``Astronomy and cosmology at Angkor Wat'', {\it Science}, 193,
(1976), 281-287.

\item Nafilyan, G. {\it Angkor Vat, Description, Graphique du
Temple}. Ecole Francaise d'Extreme-Orient, Paris. 1969.

\item Lokesh Chandra, ``Devar\={a}ja in Cambodian history'',
unpublished paper.

\item See above. Also see Lokesh Chandra, {\it Cultural Horizons
of India}. Aditya Prakashan, New Delhi, 1995.

\item C\oe d\`{e}s, G.. {\it Angkor: An Introduction.}
Oxford University Press, London, 1963, page 17.

\item Kak, S.,
``Astronomy of the Vedic altars and the Rigveda'',
{\em Mankind Quarterly}, 33, (1992), 43-55.

Kak, S., ``Astronomy in the \'{S}atapatha Br\={a}hma\d{n}a,''
{\em Indian Journal of History of Science,} 28 (1993), 15-34.

Kak, S.,
``Astronomy of the Vedic Altars,"
{\em Vistas in Astronomy,} 36 (1993), 117-140.

Kak, S., {\it The Astronomical Code of the \d{R}gveda.}
Aditya, New Delhi (1994).


Kak, S., ``The astronomy of the age of geometric altars,"
{\em Quarterly Journal of the Royal Astronomical Society,} 36 (1995) 385-396.

Kak, S., ``The sun's orbit in the Br\={a}hma\d{n}as,''
{\em Indian Journal of History of Science,} 33 (1998), 175-191.

\item
Boner, A., ``Introduction'' In 
Kaul\={a}c\={a}ra, R.,
{\it \'{S}ilpa Prak\={a}\'{s}a},
Boner, A. and Rath \'{S}arm\={a}, S. (eds.).
E.J. Brill, Leiden, 1966, pp. xxxiii.

\item
Kramrisch, S., {\it The Hindu Temple.}
The University of Calcutta, Calcutta, 1946; Motilal
Banarsidass, Delhi, 1991, page 228.

\item Kenoyer, J.M., {\it Ancient Cities of the Indus
Valley Civilization.}
Oxford University Press, Oxford, 1998, page 57.

\item See Kak, {\it The Astronomical Code ..},
pages 44-46.

\item
Kramrisch, S., {\it The Hindu Temple,} page 51.
Note that this figure is not the best modern estimate
of the period of precession.

\item
Boner, A., ``Introduction'' In 
Kaul\={a}c\={a}ra, R.,
{\it \'{S}ilpa Prak\={a}\'{s}a},
Boner, A. and Rath \'{S}arm\={a}, S. (eds.).
E.J. Brill, Leiden, 1966, pp. xxxiii.

\item SGM, page 284.

\item 
Kak, S., ``The sun's orbit in the Br\={a}hma\d{n}as,''
{\it cited above}. This altar is described in detail.
Note that a printing error caused the last sentence on
the 4th paragraph on page 187 of this paper to become
mangled. This paragraph should read: 
``If one assumes that the two halves of the year are directly in
proportional to the brick counts of 14 and 15 in the two halves
of the ring of the sun, this corresponds to day counts of
176 and 189.
This division appears to have been for the two halves of the
year with respect to the equinoxes if we note that the solstices
divide the year into counts of 181 and 184.''

\end{enumerate}

\end{document}